\documentclass[12pt,a4paper,fullpage]{article}
\usepackage{graphicx,lipsum,hyperref,url}
\usepackage[utf8]{inputenc}
\usepackage[english]{babel}
\author{Guillaume Fournier~$^\alpha$, Pierre Matoussowsky~$^\alpha$, Pascal Cotret~$^{\alpha,\beta}$\\
$^\alpha$ CentraleSupélec - Rennes, France\\
$^\beta$ IETR, SCEE Research Team - Rennes, France\\
\normalsize{Contact: \texttt{ \href{mailto:pascal.cotret@centralesupelec.fr}{pascal.cotret@centralesupelec.fr} / \href{https://twitter.com/Pascal_r2}{@Pascal\_r2}}}
}
\title{Hit the KeyJack: stealing data from your daily wireless devices incognito}
\date{}	
\begin{document}
\maketitle
\section*{Abstract}
Internet of Things (IoT) is one of the most fast-growing field in high technologies nowadays. Therefore, lots of electronic devices include wireless connections with several communication protocols (WiFi, ZigBee, Sigfox, LoRa and so on). Nevertheless, designers of such components do not take care of security features most of the time while focusing on communication reliability (speed, throughput and low power consumption). As a consequence, several wireless IoT devices transmit data in plaintext creating lots of security breaches for both eavesdropping and data injection attacks. This work introduces KeyJack, a preliminary proof-of-concept of a solution aiming to eavesdrop wireless devices and hopefully perform injection attacks afterwards. KeyJack operates on widely-used devices: our keyboards! This solution is based on low-cost embedded electronics and gives an attacker or a white hat hacker the possibility to retrieve data from John Doe's computer. This work also shows that this approach could be used to any wireless device using 2.4GHz radio chips like the NRF24L01 from Nordic Semiconductor.

\section{Introduction}
Nowadays, most of the people have IoT devices at home or even at work: for instance, it could be included in a fridge, a set-top box or computers. IoT is a highly-growing field and it will get even bigger in the next decade (a Verizon report \cite{Verizon} foresees that the IoT market will hit the \$1 trillion limit by 2019). In the same document, it can be seen that the IoT market targets several applications such as healthcare and home monitoring. Most of these connected devices were designed to perform tasks fast and in a cheap way (in other words, communication links have to be fast and low-power). Unfortunately, without security, each device is a vector of threats: malevolent people could use breaches in wireless protocols to steal or inject data using man-in-the-middle (MITM) attacks.

In this context, this work focuses on wireless keyboards that are widely-spread components. This work presents KeyJack that is a kind of wireless keylogger implemented in a tiny electronic board aiming to retrieve data from a remote computer. Furthermore, this work gives some clues about using it as a malicious injection device.

Section \ref{related} presents some related works of both classic keyloggers and embedded electronics solutions with similar goals. Section \ref{keyjack} presents KeyJack, its requirements and implementation details. Then, Section \ref{case} presents an eavesdropping scenario where Keyjack is used and gives some hints about the feasability of injection attacks using such hardware components. Finally, Section \ref{ccl} presents some conclusions and perspectives about this work.

\section{Related works}\label{related}
\subsection{Keyloggers}
When an hacker wants to retrieve data from a user keyboard, keyloggers could be used \cite{Sagiroglu,Subramayam,Holz,Dorne}: these components are usually softwares running in the background of the target computer. In more recent works such as Damopoulos et al. \cite{Damopoulos}, keyloggers are also implemented for touchscreen that are widely-spread interfaces on our smartphones and tablets. Such keyloggers aim to keep a copy of each keyboard hit made by the victim. On the other side, several works such as \cite{Herley,Ortolani} present countermeasures in order to implement systems resilient to such attacks. Therefore, IoT designers could imagine to create embedded systems immune to standard keylogger implementations.\\

However, most of software keyloggers are not so easy to use:
\begin{itemize}
	\item Advanced features are rarely included in free versions. As keyloggers may be used for "bad" purposes, developers keep the most intrusive options for paid licenses.
	\item Basic keyloggers are not 100\% discrete as they appear in the task manager.
	\item For some of them, administrator rights may be required by the operating system (for instance, installing a driver).
	\item Furthermore, results take up space on the target computer and may increase its power consumption.
\end{itemize}

\subsection{Other interesting works}
In the context of pure keyloggers, there are other interesting alternatives in the hardware community:
\begin{itemize}
	\item KeyGrabber USB made by KeeLog \cite{Keelog}.
	\item USB Rubber Ducky, a USB tool by Hak5 \cite{Rubber}. This USB key includes a 60MHz programmable microcontroller and a $\mu$SD slot. It behaves like a keyboard: therefore, nearly anything can be performed (from a Rick Roll hack to keyloggers as well).
\end{itemize}
Both solutions look like USB flash drives: it can be easily hidden on a computer port. Even if some countermeasures exists (for instance, KeyScrambler\cite{KeyScrambler} encrypts of all keyboard hits in Firefox), it is not 100\% satisfying as it still needs some physical access to the target. Furthermore, even if such a tool may be hidden in the task manager, it is assumed that its power consumption may be revealed with physical measurements.

KeyJack wants to tackle those problems using an alternative breach: nowadays, most keyboards are wireless and may be affected by eavesdropping and MITM attacks. The next subsection presents some works using wireless connection to reveal security breaches.

\subsection{Embedded electronics solutions}
There are several works aiming to steal information from wireless computer devices. The most ``industrial'' solution is MouseJack from Bastille Networks\footnote{\url{https://www.mousejack.com/}}. MouseJack is an exploit used in several wireless (non-Bluetooth) keyboards that can be used to perform eavesdropping and relay attacks. However, a laptop is required to run MouseJack in contradiction to KeyJack which aims to be a discrete and standalone solution \cite{Bastille2}.\\

Other people tried to make things smaller. Digital Security (\texttt{@iotcert}) implemented related eavesdropping methods for Bluetooth devices running on a single-board computer such as Raspberry Pi \cite{Digital1,Digital2}: Cauquil et al. implemented a Bluetooth sniffer on a RaspberryPi Zero single-board computer that can be used to track people and steal secrets (as it was shown in their Nuit du Hack'16 talk \footnote{\url{http://virtualabs.fr/ndh16/ndh16-mass-pwning-bug.pdf}}).\\

Samy Kamkar (\texttt{@samykamkar}) developed Keysweeper \cite{Kamkar1,Kamkar2}, a solution based on a small tiny microcontroller similar to Teensy or Arduino Nano. Even if this solution is the most similar to KeyJack, it does not take into account the feasability of data injection (Kamkar only focused on data listening). Furthermore, KeyJack plans to use a GSM chip to perform multiple eavesdropping with target localization in a use case where several KeyJack devices would be implemented.

\section{KeyJack}\label{keyjack}
\subsection{Threat model}
\begin{figure}[htbp]
	\centering\includegraphics[width=.55\textwidth]{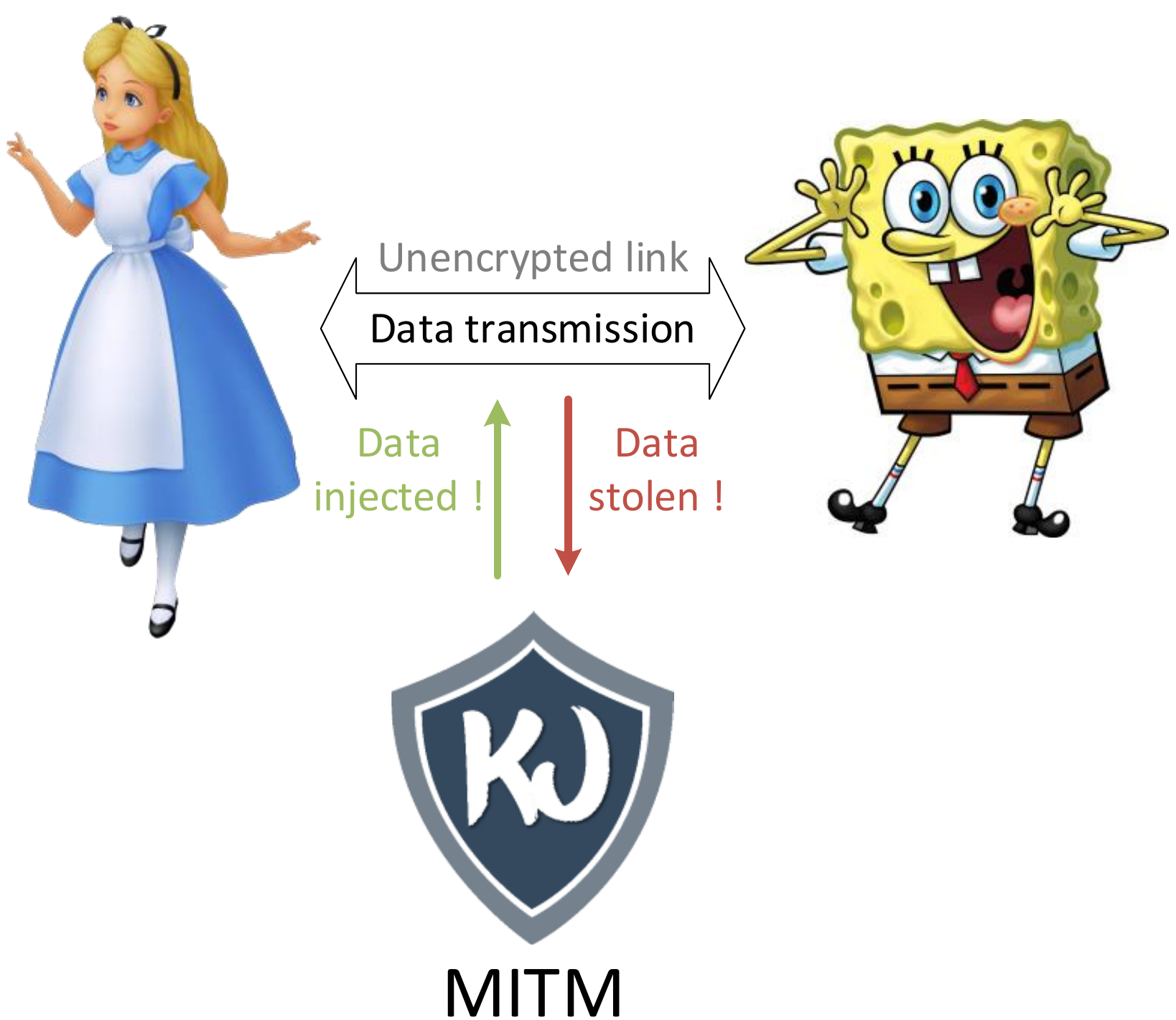}
	\caption{MITM scheme with KeyJack as the eavesdropper/injection device}
	\label{fig1}
\end{figure}

There are several keyboard manufacturers. This work focuses on the two main ones: Microsoft and Logitech (it is assumed that generic/low-cost keyboards may work as well). Microsoft/Logitech keyboards uses a classic Wifi-based protocol working at 2.4GHz (for European versions at least). When this 2.4GHz link is left unencrypted, a classic MITM scheme could be used as shown in Figure \ref{fig1}.
\\
For some recent keyboards, the communication between Alice and Bob is encrypted: even if such schemes may be broken using brute force or other advanced techniques, this work assumes that the link is left in plaintext:
\begin{itemize}
	\item For basic models, security was not implemented in the wireless protocol.
	\item For future perspectives, KeyJack could be adapted to other plaintext protocols working at different frequencies (5GHz band, for instance).
\end{itemize}

\subsection{Requirements}
A KeyJack device must have the following requirements:
\begin{itemize}
	\item No physical access to the target device/computer. It means there will not be any USB connection or malicious software installed.
	\item Tiny implementation: in other words, KeyJack must be a small-sized board, easily transportable and autonomous in terms of energy.
	\item ``Tracking-friendly'': KeyJack end-users must be able to get eavesdropping results from a remote device (website, smartphone\ldots).
	\item Injection-enabled: this work aims to propose a solution which enable not only eavesdropping but also data injection (from a remote interface as well).
	\item This work focuses on a Microsoft Wireless Keyboard 800\footnote{\url{https://www.microsoft.com/accessories/en-gb/products/keyboards/wireless-keyboard-800/2vj-00006}}. Mainly because its protocol was unecrypted.
	\item For further implementations, a GSM chip in order to retrieve locations of KeyJack nodes in case we want to install a network of such devices.
	\item And, of course, something that is low-cost and easily reproductible!
\end{itemize}
\newpage
\subsection{Hardware implementation}
KeyJack components are shown in Figure \ref{fig3}:
\begin{figure}[htbp]
	\centering\includegraphics[width=\textwidth]{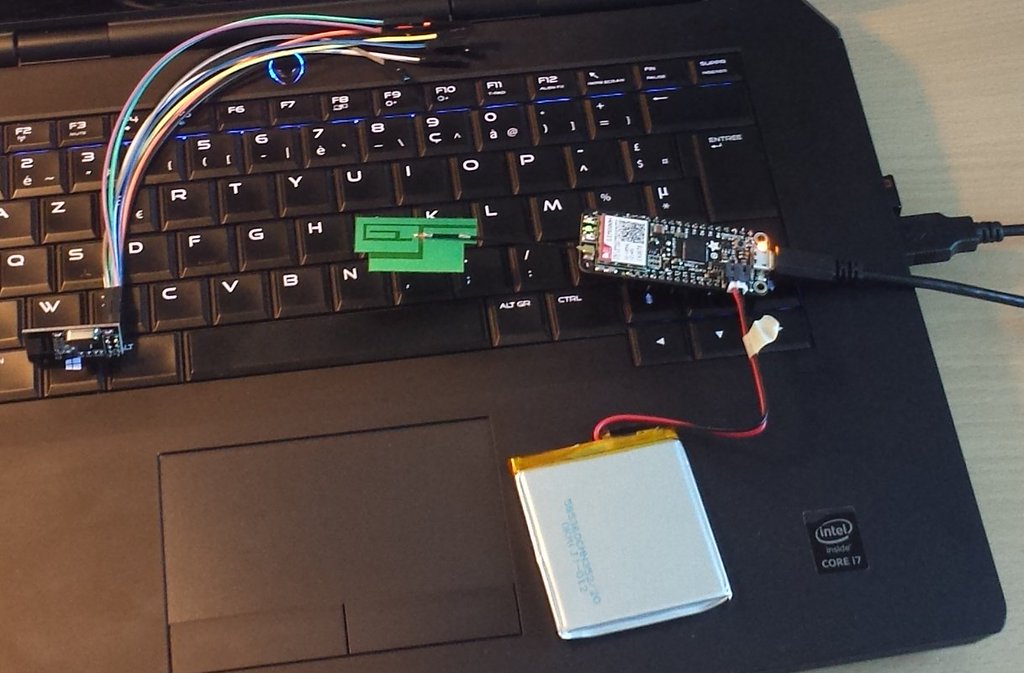}
	\caption{Electronic components used in KeyJack}
	\label{fig3}
\end{figure}~\\
From the left to the right side:
\begin{itemize}
	\item A 2.4GHz board based on a NRF24L01 chip from Nordic Semiconductors.
	\item An antenna.
	\item Adafruit Feather FONA as the microcontroller board: this board includes, in a tiny form factor, an ATmega32u4 running at 8MHz and a GSM chip.
	\item And a battery.
	\item (the USB cable is just here for programming purposes)
\end{itemize}

\subsection{Software layer}
\begin{figure}[htbp]
	\centering\includegraphics[width=.3\textwidth]{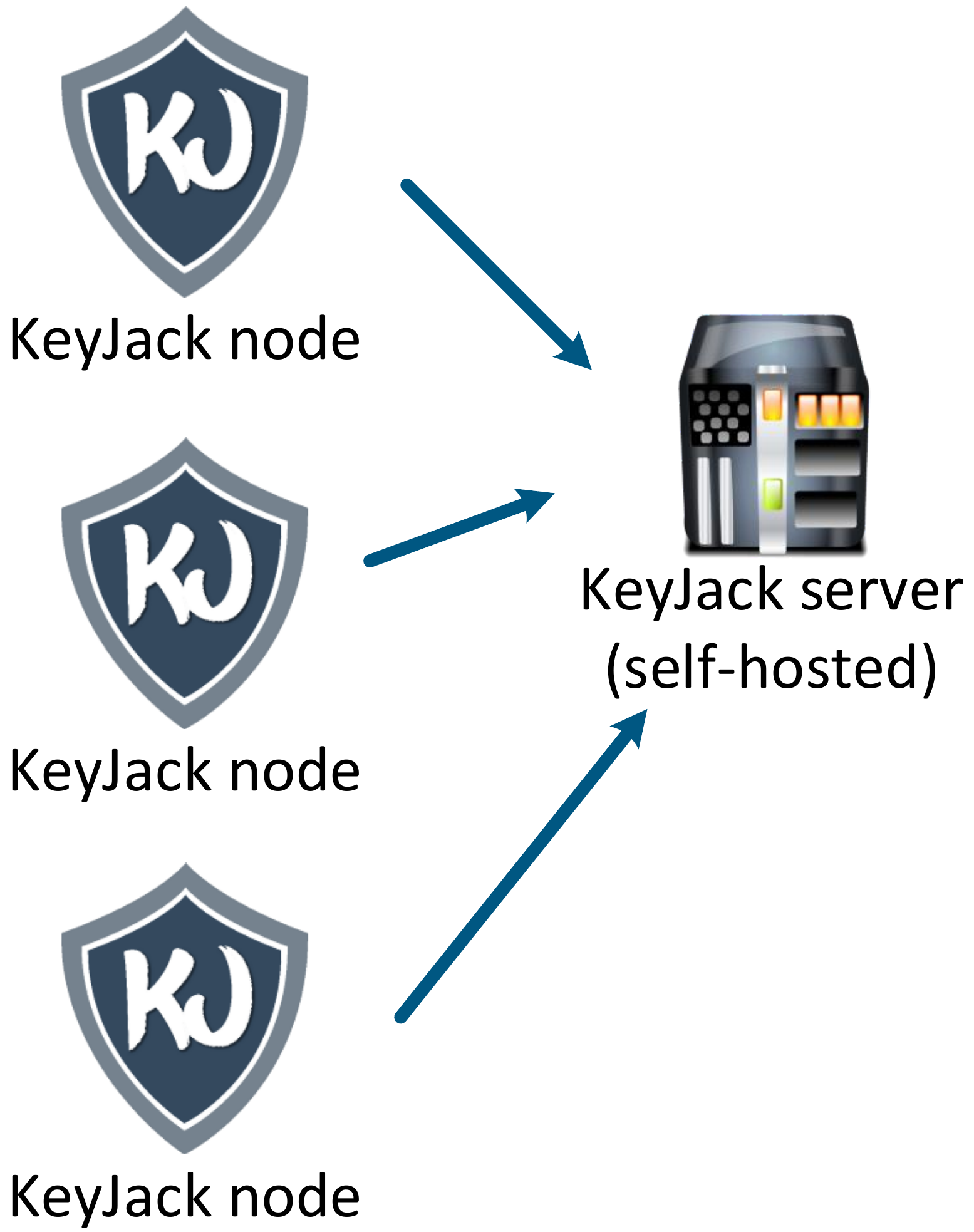}
	\caption{Example of a KeyJack network with three nodes and a server}
	\label{fig30}
\end{figure}

Figure \ref{fig30} shows an example of a KeyJack network. Each node is the Adafruit platform described in the previous section running a given Arduino code. When each node collects information, it is transmitted to the self-hosted server where an internal website was developed.
\\
On the server side, KeyJack interface looks as follows:
\begin{figure}[htbp]
	\centering\includegraphics[width=.93\textwidth]{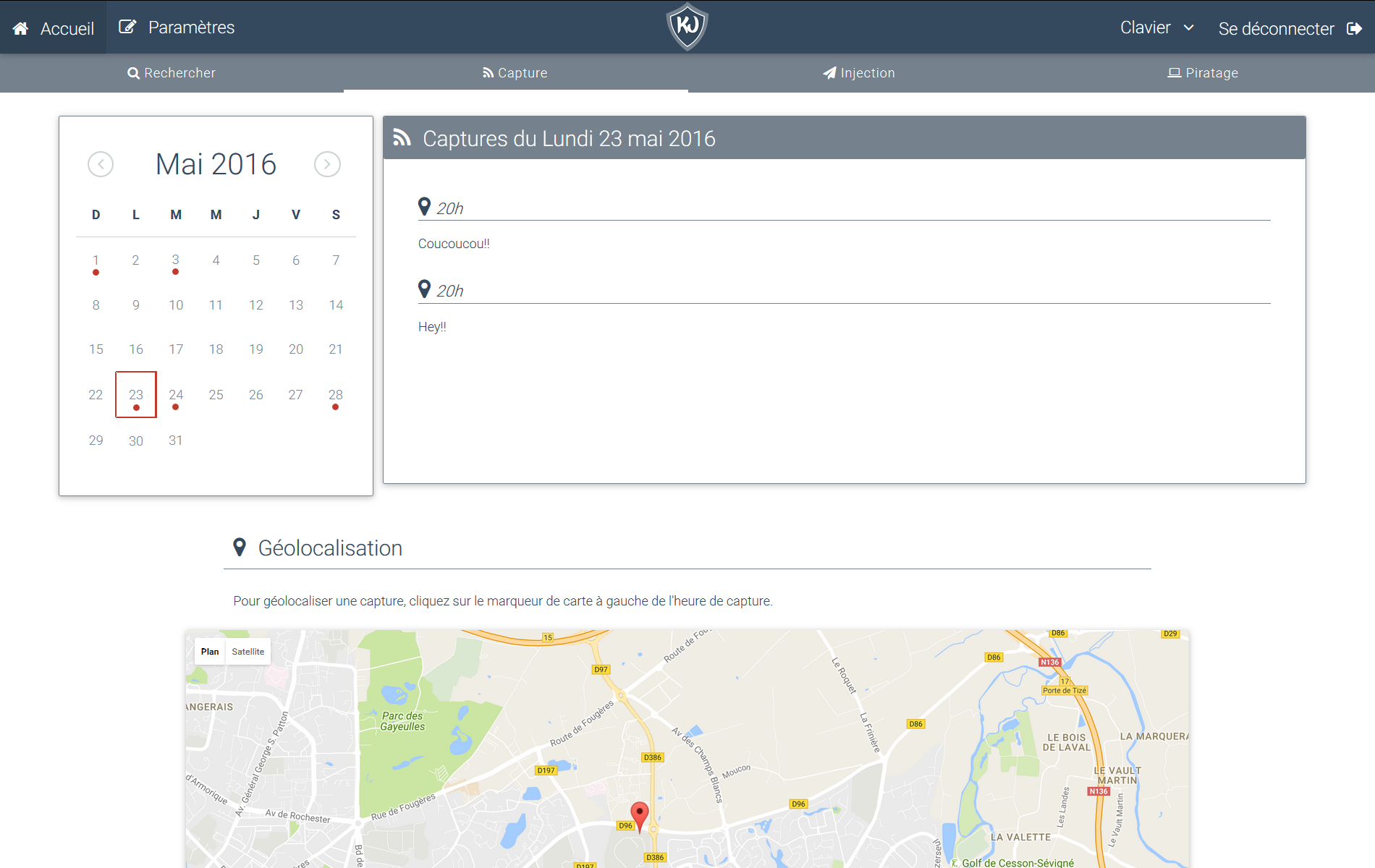}
	\caption{KeyJack server interface}
	\label{fig31}
\end{figure}

\section{Case study: eavesdropping a Microsoft keyboard}\label{case}
As it was said in Section \ref{keyjack}, this case study is focused on a Microsoft Wireless Keyboard 800. Furthermore, we only used a single KeyJack node as shown in Figure \ref{fig3}. When the user enters the KeyJack server, an interface as shown in Figure \ref{fig31} appears.
Each keyboard is identified by its MAC address and has a dedicated menu:
\begin{itemize}
	\item \textit{Search}. In this part, the user can read logs of former measurements.
	\item \textit{Capture}. Reading captures filtered by their date.
	\item \textit{Injection}. Transmitting keyboard keys.
	\item \textit{Hacking}. This last tab allows to launch attack scripts.
\end{itemize} 
\begin{figure}[htbp]
	\centering\includegraphics[width=\textwidth]{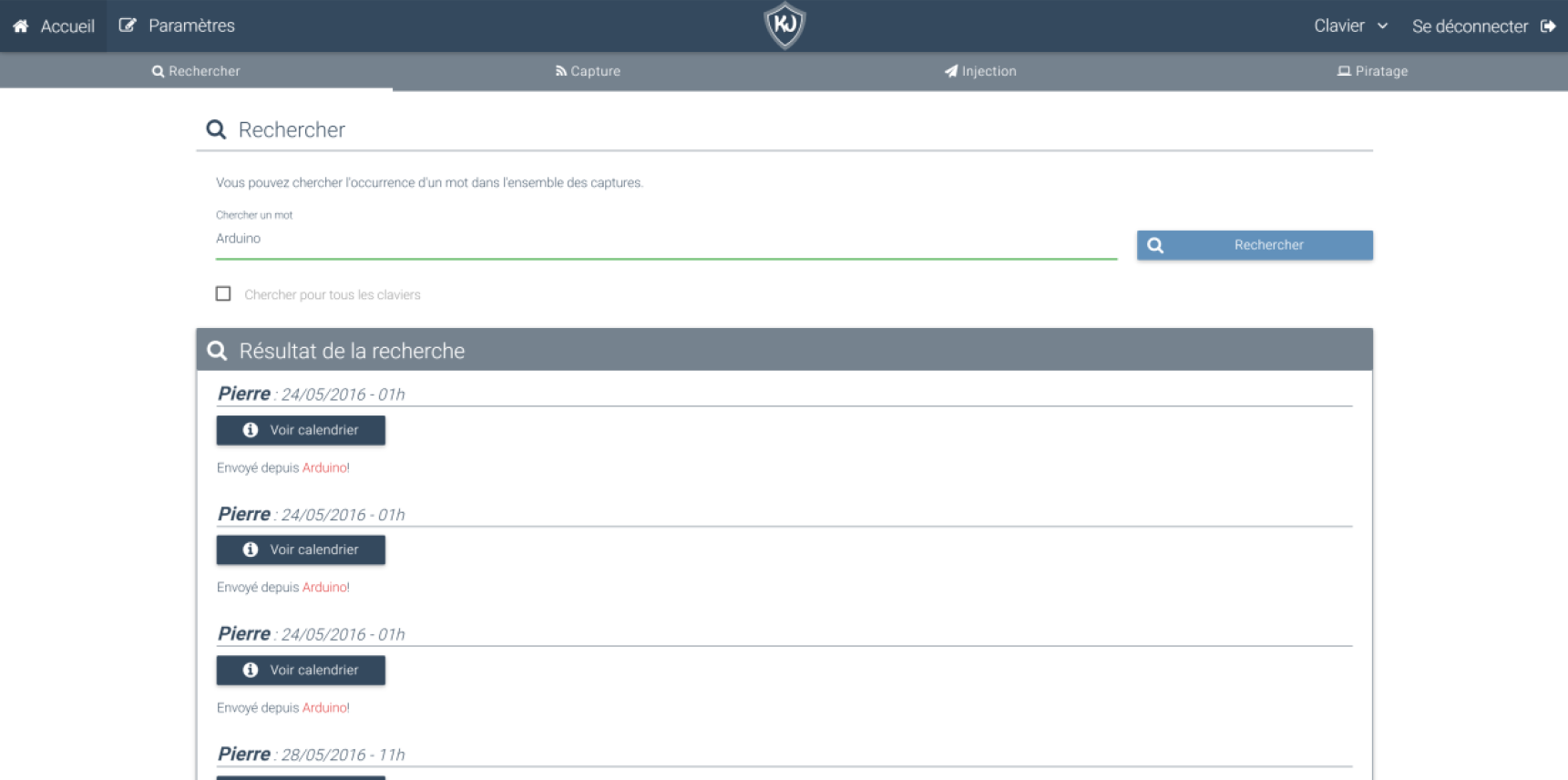}
	\caption{KeyJack keyboard interface}
	\label{fig31}
\end{figure}
Normally, the NRF24L01 chip is not able to work as a sniffer: in fact, the target MAC address is needed and it is not possible to scan the frequency spectrum around 2.4GHz to find one. However, as Samy Kamkar explained in its Keysweeper project, we can send fake information about the MAC address in order to swindle the wireless chip.\newpage
\begin{figure}[htbp]
	\centering\includegraphics[width=\textwidth]{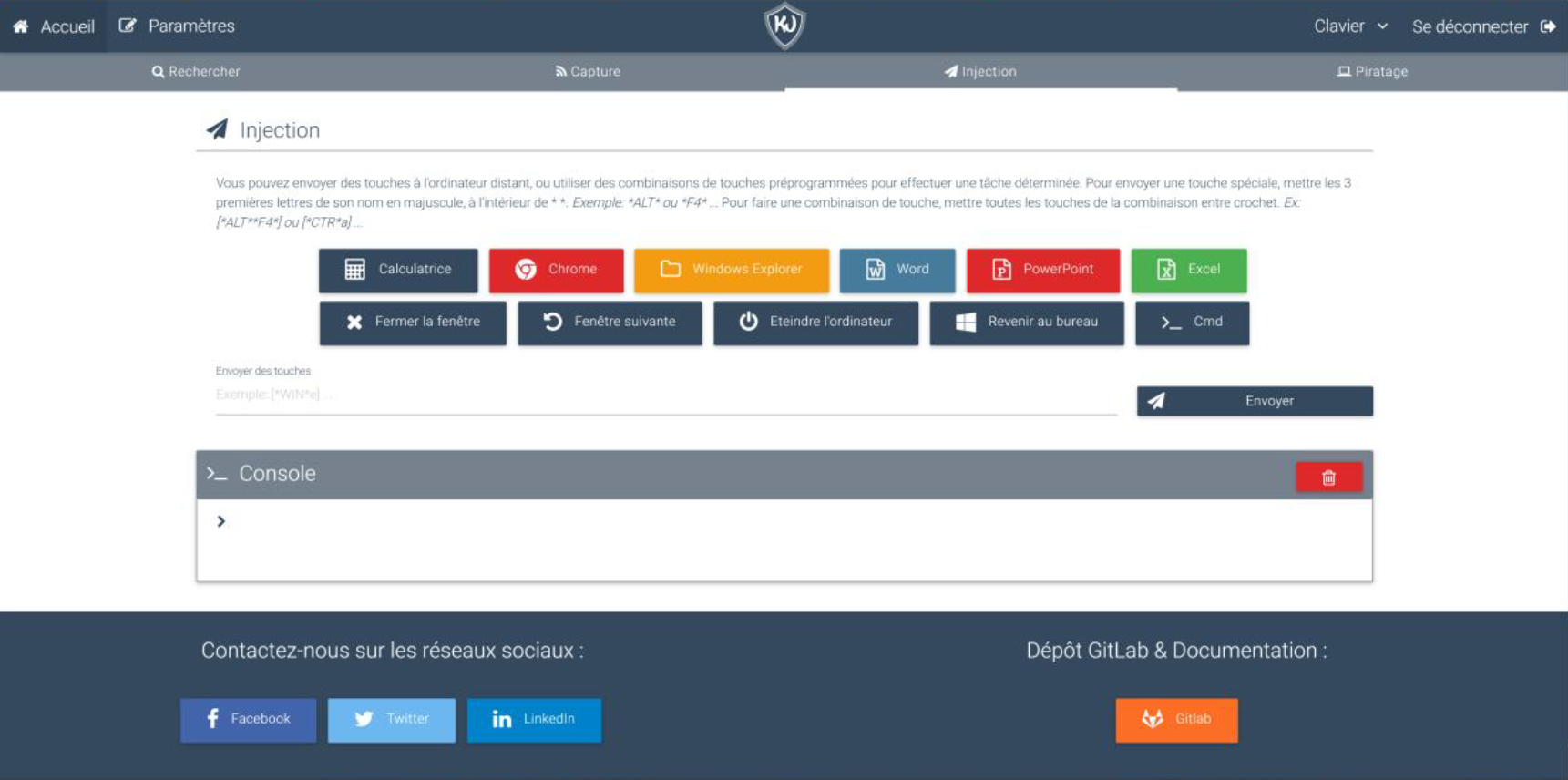}
	\caption{KeyJack injection interface}
	\label{fig31}
\end{figure}
\begin{figure}[htbp]
	\centering\includegraphics[width=\textwidth]{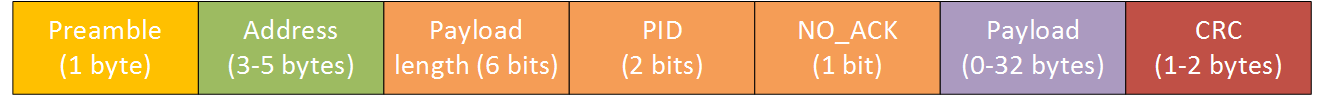}
	\caption{NRF24L01 packet structure}
	\label{fig32}
\end{figure}
With previous works of Travis Goodspeed, Samy Kamkar discovered that if we enter an incorrect value regarding the MAC address size (writing the preamble itself), the upset NRF24L01 chip considers all preambles as the target MAC address. However, all data after the MAC address should be the payload. Therfore, we get all traffic packets with the MAC address and everything else up to the CRC. From there, we can proceed to keyboard detection.\\

Each brand has its own protocol to deal with the USB dongle on the computer. Microsoft does not encrypt data sent by its keyboards (it is only done since 2015 !). The only security measure is a XOR performed on the payload with the MAC address. As we know how to get MAC address, this is not a problem for us. As a consequence, we only have to detect Microsoft protocols and perform a XOR.

\begin{figure}[htbp]
	\centering\includegraphics[width=.8\textwidth]{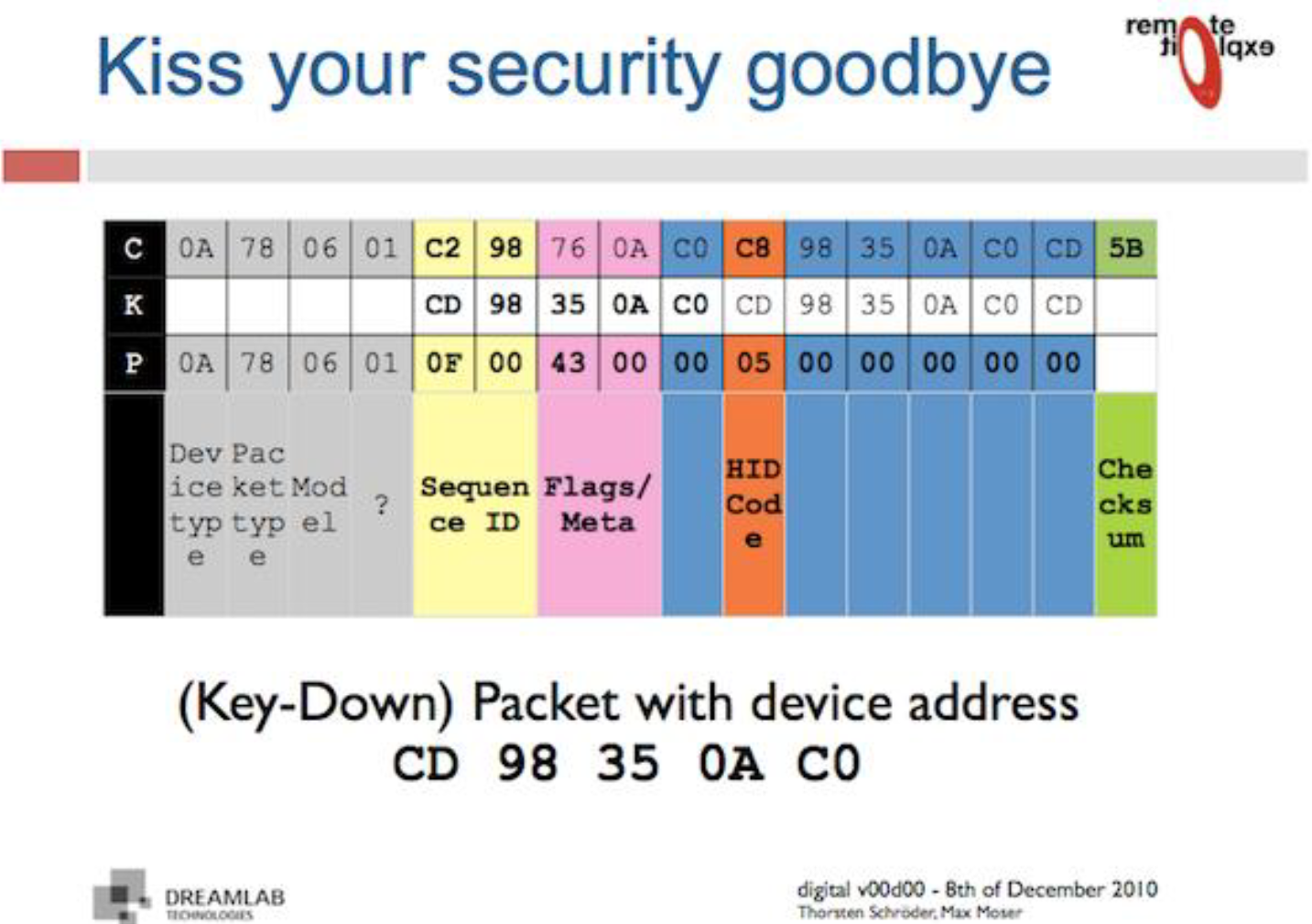}
	\caption{Packet structure}
	\label{fig42}
\end{figure}

Samy Kamkar also discovered that all Microsoft keyboards used a MAC address beginning with \texttt{0xCD} (this is the only byte we need in further measurements). In fact, Microsoft is done in a way that the key value is in the 10$^{th}$ position. As the MAC address is 5 bytes wide and 4 first bytes are not encrypted, the key value is always XORed with \texttt{0xCD}.

The last criteria we use to detect Microsoft keyboards is the value of the first unencrypted bytes. As it is shown in Figure \ref{fig42}, the device type is always \texttt{0x0A} (for keyboards). Then, the packet type indicates the key category (we only focus on keystrokes or idles). Related codes are  \texttt{0x78} and  \texttt{0x38}.
We are finally ready to scan, for all frequencies from 2403MHz to 2480MHz:
\begin{itemize}
	\item We check if the MAC address of the transmitter begins by \texttt{0xCD}.
	\item We test if the paylod begins by \texttt{0xA78} or \texttt{0xA38}.
\end{itemize}
\section{Conclusion and perspectives}\label{ccl}
This work presented KeyJack, a low-cost solution for basic eavesdropping of a specific Microsoft wireless keyboard. The proof-of-concept is a standalone device which can be left in any open-space and small enough to be hidden from people sight. Even if this study focuses on a specific keyboard model, there are opportunities with other models/vendors when communications are not encrypted. Furthermore, KeyJack can be easily modified for data injection as the server side is already implemented. As each keyboard is clearly identified, the next perspective is to make a network of KeyJack node and a single server where we could monitor everything from a remote location. Finally, KeyJack may be adapted for other protocol where security matters in the context of Internet of Things.
\bibliographystyle{plain}
\bibliography{biblio}
\end{document}